%% file: paper.tex
\documentclass[a4paper,11pt]{article}
\pdfoutput=1 

\usepackage{jcappub} 

\usepackage[T1]{fontenc} 
\usepackage{amssymb}
\usepackage{bbm}
\usepackage{txfonts}
\usepackage{psfrag}
\usepackage{graphicx}
\usepackage{natbib}
\usepackage{if then}
\usepackage{longtable}
\usepackage{ulem}
\usepackage{color}
\usepackage[dvipsnames]{xcolor}
\usepackage{tikz}
\usetikzlibrary{shapes,arrows}
\usepackage{verbatim}

\def\del#1{{}}

\renewcommand{\arraystretch}[1]{#1}

\title{Past and present cosmic structure in the SDSS DR7 main sample}

\author[a,b,1]{J. Jasche,\note{Corresponding author.}}
\author[a,b,c]{F. Leclercq,}
\author[a,b,d]{B. D. Wandelt}

\affiliation[a]{Institut d'Astrophysique de Paris (IAP), UMR 7095, CNRS - UPMC Universit\'e Paris 6, 98bis boulevard Arago, F-75014 Paris, France}
\affiliation[b]{Institut Lagrange de Paris (ILP), Sorbonne Universit\'es, 98bis boulevard Arago, F-75014 Paris, France}
\affiliation[c]{\'Ecole polytechnique ParisTech, Route de Saclay, F-91128 Palaiseau, France}
\affiliation[d]{Departments of Physics and Astronomy, University of Illinois at Urbana-Champaign, Urbana, IL 61801, USA}

\emailAdd{jasche@iap.fr}
\emailAdd{florent.leclercq@polytechnique.org}
\emailAdd{wandelt@iap.fr}

\abstract{
We present a chrono-cosmography project, aiming at the inference of the four dimensional formation history of the observed large scale structure from its origin to the present epoch. To do so, we perform a full-scale Bayesian analysis of the northern galactic cap of the Sloan Digital Sky Survey (SDSS) Data Release 7 main galaxy sample, relying on a fully probabilistic, physical model of the non-linearly evolved density field. Besides inferring initial conditions from observations, our methodology naturally and accurately reconstructs non-linear features at the present epoch, such as walls and filaments, corresponding to high-order correlation functions generated by late-time structure formation. Our inference framework self-consistently accounts for typical observational systematic and statistical uncertainties such as noise, survey geometry and selection effects. We further account for luminosity dependent galaxy biases and automatic noise calibration within a fully Bayesian approach. As a result, this analysis provides highly-detailed and accurate reconstructions of the present density field on scales larger than $\sim~3$ Mpc$/h$, constrained by SDSS observations. This approach also leads to the first quantitative inference of plausible formation histories of the dynamic large scale structure underlying the observed galaxy distribution. The results described in this work constitute the first full Bayesian non-linear analysis of the cosmic large scale structure with the demonstrated capability of uncertainty quantification. Some of these results will be made publicly available along with this work. The level of detail of inferred results and the high degree of control on observational uncertainties pave the path towards high precision chrono-cosmography, the subject of simultaneously studying the dynamics and the morphology of the inhomogeneous Universe.}


\begin{document}
\date{Accepted 20?? December ??.
      Accepted 20?? December ??;
      in original form 20?? October ??}


\maketitle
\flushbottom

\section{Introduction}
Analyzing the three dimensional (3D) cosmic large scale structure (LSS), has been a major endeavor in modern cosmology for several decades now. Novel and progressively more sensitive analysis methods are continuously developed and applied to cosmological surveys in order to answer outstanding questions on the origin and evolution of our Universe. Gravitational forces, sourced by well known baryonic matter and radiation but also by enigmatic dark matter and dark energy, shaped the presently observed cosmic large scale structure via non-linear clustering of primordial density fluctuations. As a consequence, the distribution of matter in our Universe contains a wealth of information on governing physical and dynamic structure formation processes. Observations of probes of this cosmic matter distribution, such as galaxies, can therefore act as powerful laboratories to confirm our current understanding of cosmology and test fundamental physics if non-trivial connections between theory and observations can be established. In particular, cosmographic descriptions of the LSS, in terms of 3D maps of the matter distribution, reveal the details of gravitational structure formation, the clustering behavior of galaxies and permit to characterize initial conditions and large scale cosmic flows.

\begin{figure*}
	\centering{\includegraphics[width=0.66\textwidth,clip=true]{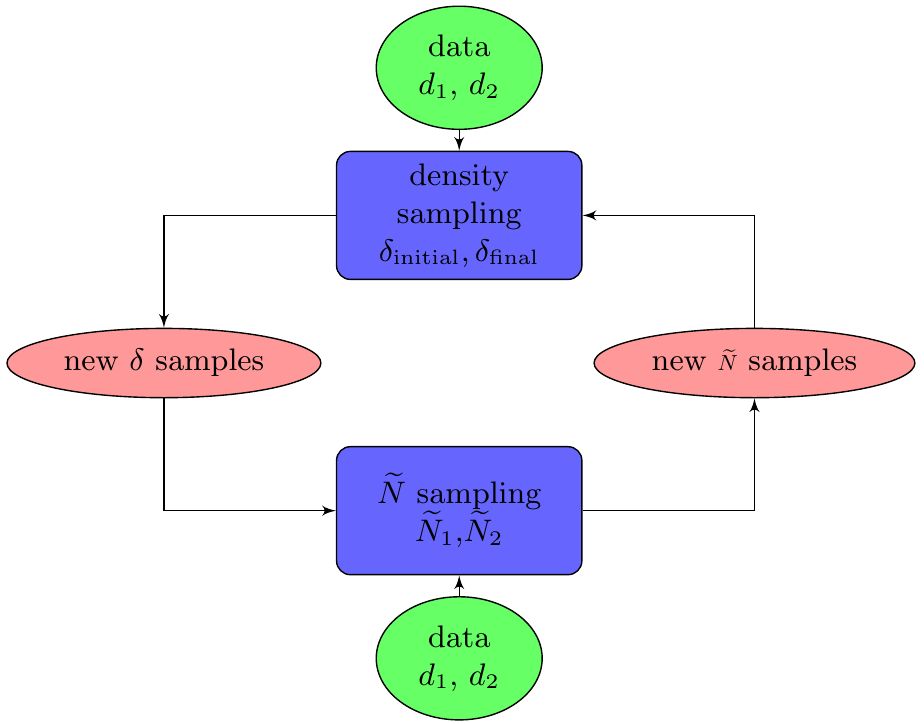}}
\caption{Flow chart depicting the multi-step iterative block sampling procedure exemplified for two data sets. In the first step, the {\bf{BORG}} algorithm sampler generates realizations of the initial and final density fields conditioned to the galaxy samples $d_1$ and $d_2$. In an subsequent step, the noise parameters \(\widetilde{N}\) will be sampled conditional on the respective previous density realizations.}
	\label{fig:flowchart}
\end{figure*}
For these reasons, inferred 3D density and velocity maps have wide spread applications in modern cosmology. In particular, there exists a rich literature on inferring cosmological power spectra from the 3D LSS \citep[see e.g.][]{FELDMAN1994,TEGMARK1995,BALLINGER1995,HEAVENS1995,HAMILTON1997A,TEGMARK1997,TADROS1999,YAMAMOTO2003,
TEGMARK_2004,COLE2005,PERCIVAL2005,2006A&A...459..375H,JASCHESPEC2010,JaschePspec2013}. We note that in principle, the availability of inferred 3D density fields permits to extend clustering analyses beyond two- to three-point and higher-order statistics. Reconstructions of 3D density fields are also used for generating templates of the integrated Sachs-Wolfe (ISW) or kinetic Sunyaev-Zeldovich (kSZ) effects in cosmic microwave background observations \citep[see e.g.][]{2008MNRAS.391.1315F,2008PhRvD..77l3520G,2008PhRvD..78d3519H,2009ApJ...701..414G,2013arXiv1303.5079P,2013MNRAS.430..259B,2014MNRAS.438.1724H,Li2014}. In particular, Li et al. recently proposed a optimal matched filter approach, relying on density field reconstructions, to detect the kSZ signal \citep{Li2014}. In an analogous fashion, reconstructions of the 3D density field permit to build templates for weak lensing signals \citep[see e.g.][]{2012MNRAS.424..553A,2014MNRAS.440.2191S}. Cosmological constraints can also be derived from the topology of the 3D matter distribution. In particular, Minkowski functionals, genus statistics or the skeleton of the cosmic large scale structure have been discussed in the literature \citep[][]{1994A&A...288..697M,1994ApJ...420..525V,1999ApJ...526..568S,2002PASJ...54..707H,2003PASJ...55..911H,2009MNRAS.393..457S,2010ApJ...723..364A,2011MNRAS.414..384S,2012MNRAS.423.3209P}. 
Various methods aiming at classifying different LSS objects, such as filaments, voids and clusters, in Lagrangian or Eulerian density and velocity fields, have been developed and applied to observations \citep[see e.g.][]{HAHN2007,FORERO2009,JASCHE2010HADESDATA,2010MNRAS.409..156B,2010MNRAS.403.1392L,2012MNRAS.425.2049H,2013JCAP...11..048L,
2013arXiv1309.4787N,2013MNRAS.428..141N,2013MNRAS.429.1286C,2014arXiv1406.1191S,2014arXiv1406.1004N}. 

A particularly interesting approach to test the origin and evolution of our Universe uses phase information of three dimensional density fields \citep[see e.g.][]{2000Natur.406..376C,2002MNRAS.337..488C,2003ApJ...589L..61W,2004ApJ...600..553H,2005PASJ...57..709H,2013ApJ...762..115O}.
In particular, Hikage et al. pointed out that measures of triangular sums of phase angles are related to measurements of the cosmic bi-spectrum \citep{2004ApJ...600..553H}. Furthermore, Obreschkow et al. showed that correlation functions of the phase distribution in 3D density fields may act as a powerful probe to test cosmology and to infer the mass of warm dark matter particles \citep{2013ApJ...762..115O}.

Beyond immediate cosmological applications, it is also known that physical properties of galaxies and other cosmological observables are related to their cosmic large scale environments. For these reasons, reconstructed density fields are also traditionally used to test properties such as morphological type, color, luminosity, spin
parameter, star formation rate, concentration parameter, etc., as
functions of their cosmic environments \citep[see
e.g.][]{DRESSLER1980,POSTMAN1984,WHITMORE1993,LEWIS2002,GOMEZ2003,GOTO2003,
ROJAS2005,KUEHN2005,BLANTON2005,BERNARDI2006,CHOI2007,PARK2007,LEE2008,LEELI2008,2014MNRAS.438..717K}.

Already this incomplete list of possible applications demonstrates the demand for accurate 3D large scale structure reconstructions. 
As a response to this demand, a variety of methods to infer density and velocity fields from observations, have been developed and applied to data \citep[see e.g.][]{1993PhRvE..47..704E,HOFFMAN1994,1994ASPC...67..171L,1994ApJ...423L..93L,1995A&AS..109...71Z,
1995MNRAS.272..885F,1995ApJ...449..446Z,1997MNRAS.287..425W,1999ApJ...520..413Z,
2001misk.conf..268V,2006MNRAS.373...45E,ERDOGDU2004,2008PhyD..237.2145M,KITAURA2009MNRAS,KITAURA2010MNRAS,2010ApJ...708..505K,JASCHE2010HADESMETHOD,
JASCHESPEC2010,2011MNRAS.417.1303M,JASCHE2012PHOTOZ,JaschePspec2013,JASCHEBORG2012,2013MNRAS.429L..84K,2013ApJ...772...63W}.
Further scientific interpretation of these reconstructions requires an understanding of the uncertainties inherent in these analyses and a means to propagate these to any finally inferred quantity in order not to draw erroneous conclusion on the theoretical model to be tested. Three-dimensional inference of the matter distribution from observations requires modeling the statistical behavior of the mildly non-linear and non-linear regime of the matter distribution. The exact statistical behavior of the matter distribution in terms of a multivariate probability distribution for fully gravitationally evolved density fields is not known. Previous approaches therefore typically relied on phenomenological approximations such as multivariate Gaussian or log-normal distributions incorporating a cosmological power spectrum to accurately account for the correct two-point statistics of density fields \citep[see e.g.][and references therein]{HOFFMAN1994,1994ApJ...423L..93L,1995MNRAS.272..885F,2006MNRAS.373...45E,KITAURA2009MNRAS,KITAURA2010MNRAS,JASCHESPEC2010,JASCHE2010HADESMETHOD,JASCHE2010HADESDATA,JaschePspec2013}. Both of these distributions can be considered as maximum entropy prior on linear and logarithmic scales, respectively, and are well-justified for Bayesian analyses. However, these priors only approximate the statistical behaviour of density fields up to two-point statistics. 

As pointed out by Jasche and Wandelt in \citep[][]{JASCHEBORG2012}, large scale structure formation through gravitational clustering is essentially a deterministic process described by Einstein's equations and since the only stochasticity in the problem enters in the generation of initial conditions, it seems reasonable to account for the increasing statistical complexity of the evolving matter distribution by a dynamical model of structure formation. This approach of using data to constrain a set of \textit{a priori} possible dynamical, three-dimensional structure formation histories has been recently proposed and implemented into the Bayesian inference framework {\bf{BORG}} (Bayesian Origin Reconstruction from Galaxies, \citep{JASCHEBORG2012}). This methodology employs second order Lagrangian perturbation theory (2LPT) as a physical model of the gravitational dynamics linking the initial 3D Gaussian density field to the presently observed, non-Gaussian density field. In particular, this algorithm uses powerful sampling techniques to correctly explore the range of possible initial density fields that are statistically consistent with present observations, modeled as a Poisson sample from evolved density fields. This method accurately infers three dimensional initial and final density fields as well as large scale velocity fields and provides corresponding uncertainty quantification in a fully Bayesian framework. In particular, the variations between different data-constrained realizations, generated by the sampling algorithm, represent the uncertainties that remain in the reconstruction owing to the modeled statistical and systematic errors in the data. A natural and particularly interesting byproduct of this methodology is the ability to infer plausible structure formation histories, describing the evolution of the observed LSS from its origins to the present epoch. Therefore the algorithm not only provides highly detailed and accurate reconstructions of 3D density and velocity fields, but it also provides entire data-constrained reconstructions of the four dimensional (4D) dynamic and evolving state of our Universe. Together with accurate uncertainty quantification, as provided by the {\bf{BORG}} algorithm, this feature enables us to progress towards high precision chrono-cosmography, the subject of inferring the formation history of our Universe.

This work focuses on the description of such a chrono-chosmography project conducted with real data. We apply the {\bf{BORG}} algorithm to the Sloan Digital Sky Survey DR7 main galaxy sample \citep[][]{YORK2000,SDSS7}. The primary intention of this work is to provide accurate reconstructions of the 3D gravitationally evolved density field, consisting of filaments, voids and clusters along with corresponding uncertainty quantification. In doing so, our algorithm will automatically and fully self-consistently provide inferences for corresponding initial conditions at a cosmic scale factor of \(a=10^{-3}\) and 3D velocity fields at the present epoch. Further, we will be able to provide plausible, data-constrained full 4D formation histories for the cosmic large scale structure inside the domain covered by SDSS observations.

Besides typical systematic and stochastic uncertainties, such as survey geometry, selection effects and noise, the algorithm also accounts for luminosity dependent galaxy bias of DR7 main sample galaxy populations. To allow for different luminosity dependent galaxy biases in the data model, we followed the approach described in \citep{JaschePspec2013} and upgraded the original formulation of the {\bf{BORG}} algorithm correspondingly. This modification permits us to sub-divide the galaxy sample into different sub-samples according to their respective luminosities. Each of these sub-samples can be treated individually with respect to galaxy biases, selection effects and noise levels. To automatically calibrate corresponding noise levels, we introduced an additional sampling block to the original {\bf{BORG}} sampling framework, as outlined in figure \ref{fig:flowchart}. As a result, the upgraded version of the {\bf{BORG}} algorithm permits to jointly infer 3D LSS fields and noise levels inherent to SDSS observations.

The present work describes the first full Bayesian non-linear analysis of the cosmic LSS with the demonstrated capability of quantifying uncertainties in a high dimensional setting. 
Specifically the application of the {\bf{BORG}} algorithm to the SDSS DR7 main galaxy sample yields highly detailed and accurate reconstructions of 3D density and velocity fields even in regions that are only poorly sampled by observations. Together, with a thorough uncertainty quantification, as provided by our methodology, these results form the basis for several upcoming publications which aim at analysing obtained results in more detail and on a more quantitative level. In particular an upcoming publication will focus on the analysis of dark matter voids in the SDSS main galaxy sample \citep[][]{Leclercq2014A}. 
We will also make some of our results publicly available. In particular, the posterior mean final density field together with corresponding standard deviations will be  published as supplementary material along with this article.\footnote{Prior to the publication at JCAP please contact us to receive a copy of this data.}

This work is structured as follows. In section \ref{sec:galaxy_sample} we give a brief overview about the SDSS data set used in this work, followed by a description of our Bayesian inference methodology in section \ref{sec:methodology}. We also discuss modifications of the original algorithm, previously described in \citep{JASCHEBORG2012}, to account for luminosity dependent galaxy bias and automatic noise calibration. In section \ref{sec:Bayesian_inference}, we demonstrate the application of our inference algorithm to observations and discuss the general performance of the Hamiltonian Monte Carlo sampler. Section \ref{sec:inference_results} describes the inference results obtained in the course of this project. In particular, we present results on inferred 3D initial and final density as well as velocity fields and show the ability of our method to provide accurate uncertainty quantification for any finally inferred quantity. Further, we will also demonstrate the ability of our methodology to perform chrono-cosmography, by accurately inferring plausible 4D formation histories for the observed LSS from its origins to the present epoch. We conclude this paper in section \ref{sec:Summary_an_Conclusion}, by summarizing and discussing the results obtained in the course of this work.

\section{The SDSS galaxy sample}
\label{sec:galaxy_sample}
In this work, we follow a similar procedure as described in \citep{JASCHE2010HADESDATA}, by applying the {\bf{BORG}} algorithm to the SDSS main galaxy sample. Specifically, we employ the {\tt Sample dr72} of the New York University Value Added Catalogue (NYU-VAGC).\footnote{http://sdss.physics.nyu.edu/vagc/} This is an updated version of the catalogue originally constructed by Blanton et al. \citep{BLANTON2005} and is based on the final data release \citep[DR7;][]{SDSS7} of the Sloan Digital Sky Survey \citep[SDSS;][]{YORK2000}. Based on {\tt Sample dr72}, we construct a flux-limited galaxy sample with spectroscopically measured redshifts in the range $0.001<z<0.4$, $r$-band Petrosian apparent magnitude $r\leq 17.6$ after correction for Galactic extinction, and $r$-band absolute magnitude $-21<M_{^{0.1}r}<-17$. Absolute $r$-band magnitudes are corrected to their $z=0.1$ values using the $K$-correction code of Blanton et al. \citep{BLANTON2003A,BLANTON2007}
and the luminosity evolution model described in \citep{BLANTON2003}. We also restrict our analysis to the main contiguous region of the SDSS in the northern Galactic cap, excluding the three survey strips in the southern cap (about 10 per cent of the full survey area). The NYU-VAGC provides required information on the
incompleteness in our spectroscopic sample. This includes a mask, indicating which areas of the sky have been targeted and which not. The mask defines the effective area of the survey on the sky, which is 6437 deg$^2$ for the sample we use here. This survey area is divided into a large number of smaller subareas, called {\it polygons}, for each of which the NYU-VAGC lists a spectroscopic completeness, defined as the fraction of photometrically identified target galaxies in the polygon for which usable spectra were obtained. Throughout our sample the average completeness is 0.92. To account for radial selection functions, defined as the fraction of galaxies in the absolute magnitude range considered here, that are within the apparent magnitude range of the sample at a given redshift, we use a standard luminosity function proposed by Schechter in \citep{SCHECHTER1976} with $r$-band parameters ($\alpha = -1.05$, $M_{*} - 5\mathrm{log}_{10}(h) = -20.44$) \citep[][]{2003ApJ...592..819B}.

We note that our analysis accounts for luminosity dependent galaxy biases, by following a similar approach to the one described in \citep{JaschePspec2013}. In order to do so we subdivide our galaxy sample into six equidistant bins in absolute $r$-band magnitude in the range $-21<M_{^{0.1}r}<-17$, resulting in a total of $372,198$ main sample galaxies to be used in the analysis. 
 As will be described in more detail below, splitting the galaxy sample permits us to treat each of these sub-samples as an individual data set, with its respective selection effects, biases and noise levels.    

\begin{figure*}
\centering{\includegraphics[width=1.0\textwidth,clip=true]{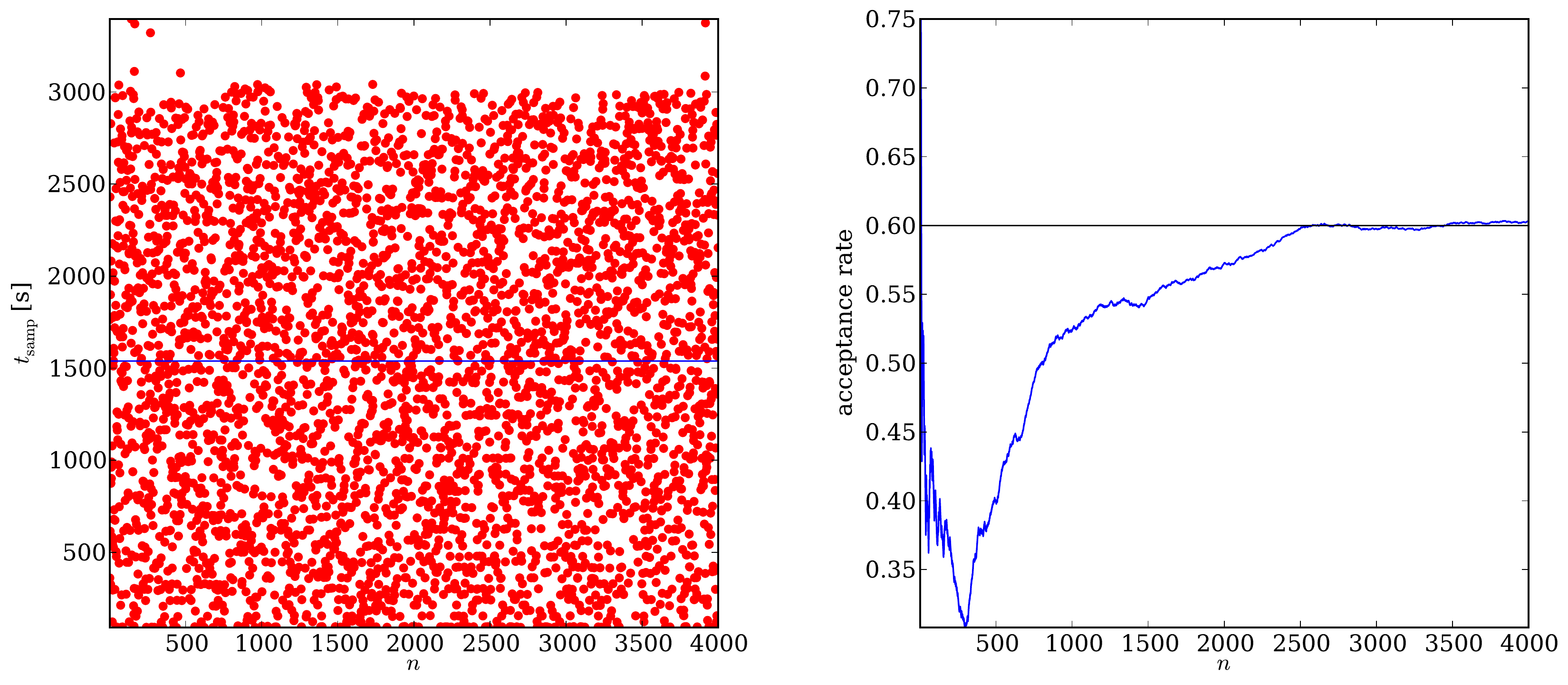}}
\caption{The scatter of sample generation times (left panel) and Markov acceptance rates during the initial burn-in phase (right panel). As shown by the left panel, times to generate individual samples range from zero to about 3000 seconds. The average execution time per sample generation is about 1500 seconds on \(16\) cores. Initially, acceptance rates drop during burn-in but rise again to reach an asymptotic value of about 60 percent.}
\label{fig:markov_properties}
\end{figure*}

\section{Methodology}
\label{sec:methodology}

In this section, we give a brief introduction to the Bayesian inference framework {\bf{BORG}} (Bayesian Origin Reconstruction from Galaxies) and describe some minor updates to the original algorithm described in \citep{JASCHEBORG2012}.
 
\subsection{The Bayesian inference framework}
This work presents an application of the {\bf{BORG}} (Bayesian Origin Reconstruction from Galaxies, \citep{JASCHEBORG2012}) algorithm to SDSS main galaxies. The {\bf{BORG}} algorithm is a full scale Bayesian inference framework aiming at the analysis of the linear and mildly non-linear regime of the cosmic LSS. It performs dynamical large scale structure inference from galaxy redshift surveys employing a physical model for gravitational structure formation. Introduction of physical models to the inference process generally turns large scale structure inference into the task of inferring corresponding initial conditions at an earlier epoch from present cosmological observations. The resultant procedure is numerically highly non-trivial, since it requires to explore the very high-dimensional and non-linear space of possible solutions to the initial conditions problem within a fully probabilistic approach. Typically, these spaces comprise $10^6$ to $10^7$ parameters, corresponding to the voxels used to discretize the observed domain.

The {\bf{BORG}} algorithm explores a large scale structure posterior consisting of a Gaussian prior for the initial density field at a initial cosmic scale factor of $a=10^{-3}$ linked to a Poissonian model of galaxy formation at a scale factor $a=1$ via a second order Lagrangian perturbation theory model \citep[for details see][]{JASCHEBORG2012}. As described in the literature, 2LPT describes the one, two and three-point statistics correctly on scales larger than \(\sim 8 \, \mathrm{Mpc}/h\) and represents higher-order statistics very well \citep[see e.g. ][]{MOUTARDE1991,BUCHERT1994,BOUCHET1995,SCOCCIMARRO2000,PTHALOS, 2013JCAP...11..048L}. The {\bf{BORG}} algorithm therefore naturally accounts for the filamentary structure of the cosmic web typically associated to higher-order statistics induced by non-linear gravitational structure formation processes. The posterior distribution also accounts for systematic and stochastic uncertainties, such as survey geometries, selection effects and noise typically encountered in cosmological surveys.

Numerically efficient exploration of this highly non-Gaussian and non-linear posterior distribution is achieved via an efficient implementation of a Hamiltonian Markov Chain Monte Carlo sampling algorithm \citep[see][for details]{DUANE1987,JASCHEBORG2012}. It is important to remark that our inference process requires at no point the inversion of the flow of time in the dynamical model. The analysis solely depends on forward evaluations of the model, alleviating many of the problems encountered in previous approaches to the inference of initial conditions, such as spurious decaying mode amplification \citep[see e.g.][]{NUSSER1992,Crocce2006}. Rather than inferring the initial conditions by backward integration in time, our approach builds a fully probabilistic non-linear filter using the dynamical forward model as a prior. This prior singles out physically reasonable large scale structure states from the space of all possible solutions. Since the {\bf{BORG}} algorithm provides an approximation to non-linear large scale dynamics, it automatically provides information on the dynamical evolution of the large scale matter distribution. In particular, it explores the space of dynamical structure formation \textit{histories} compatible with both data and model. Also note, that the {\bf{BORG}} algorithm naturally infers initial density fields at their Lagrangian coordinates, while final density fields are recovered at corresponding final Eulerian coordinates. Therefore the algorithm accounts for the displacement of matter in the course of structure formation. In the present algorithm redshift space distortions have not been accounted for explicitly. However, according to its current formulation the {\bf{BORG}} algorithm interprets features associated to redshift distortions as noise and will tend to infer isotropic density fields. Isotropy of density fields is naturally imposed by assuming diagonal covariance matrices for initial density fields in the prior distribution, which reflects the cosmological principle.
As a consequence, this algorithm not only provides measurements of the 3D density field but also performs full four-dimensional state inference and recovers the dynamic formation history of the large scale structure, opening the possibility to do 4D chrono-cosmography.

\subsection{Updating the data model}

The data model employed in this work assumes a Poissonian picture of galaxy formation to account for the noise of a discrete galaxy distribution \citep[for details see e.g. ][]{MARTINEZ2002,JASCHE2010HADESDATA,JASCHE2010HADESMETHOD,JASCHEBORG2012}. In addition to the original data model, described in \citep{JASCHEBORG2012}, we also account for luminosity dependent galaxy bias in a similar spirit as described in \citep{JaschePspec2013}. By sub-dividing the galaxy sample into several different bins of absolute $r$-band magnitude, our approach permits to account for respective uncertainties, systematics and galaxy biases of each individual sub-sample. While Jasche and Wandelt assumed a linear bias model in \citep{JaschePspec2013}, here we have to extend this approach to a non-linear bias model. Indeed, in order to be well defined, a Poisson likelihood requires the intensity for inhomogeneous Poisson processes to be strictly positive. Since a linear bias does not guarantee a positive density field and corresponding Poisson intensity, this bias model is not applicable to the present case. For this reason, we assume a power-law model to account for galaxy biasing \citep[][]{1998MNRAS.293..209M,2000ApJ...528....1N}. In this picture, the relation between the matter density field $\rho_\mathrm{m}$ and the expected galaxy density field $\rho_\mathrm{g}^\ell$ for galaxies in the $\ell$th absolute magnitude bin can be described by:
\begin{equation}
\rho_\mathrm{g}^\ell = \beta^\ell \, \rho_\mathrm{m}^{\alpha^\ell} = \beta^\ell \, \left(1+\delta\right)^{\alpha^\ell} ,
\end{equation}
where $\alpha^\ell$ and $\beta^\ell$ are bias parameters for the $\ell$th absolute magnitude bin and $\delta$ is the density contrast of the matter field. The Poisson likelihood for observed galaxy number counts $N_i^\ell$ can then be written as:
\begin{equation}
\label{eq:poisson_likelihood}
\mathcal{P}\left( \{ N^\ell_i \} \vert \{ \delta_i \} \right) = \prod_i \mathrm{e}^{-\bar{N}^\ell \, R_i^\ell \, \beta^\ell \left( 1+ \delta_i \right)^{\alpha^\ell}} \, \frac{\left(\bar{N}^\ell \, R_i^\ell \, \beta^\ell \left( 1+ \delta_i \right)^{\alpha^\ell} \right)^{N^\ell_i}}{N^\ell_i!} ,
\end{equation}
where $i$ labels a voxel, $R_i^\ell$ is the survey response operator consisting of the product of the radial selection function and the angular survey mask and completeness function, $\bar{N}^\ell$ is the expected mean number of galaxies, setting the expected noise level in the survey \citep[for details on the Poisson likelihood see e.g.][]{JASCHE2010HADESDATA,JASCHE2010HADESMETHOD,JASCHEBORG2012}. As can be seen from equation (\ref{eq:poisson_likelihood}), the parameters $\bar{N}^\ell$ and $\beta^\ell$ are degenerate and can be replaced by the product quantity $\widetilde{N}^\ell=\bar{N}^\ell\,\beta^\ell$. The resulting likelihood distribution reads:
\begin{equation}
\label{eq:poisson_likelihood_2}
\mathcal{P}\left( \{ N^\ell_i \} \vert \{ \delta_i \} \right) = \prod_i \mathrm{e}^{-\widetilde{N}^\ell \, R_i^\ell \, \left( 1+ \delta_i \right)^{\alpha^\ell}} \, \frac{\left(\widetilde{N}^\ell \, R_i^\ell \, \left( 1+ \delta_i \right)^{\alpha^\ell}\right)^{N^\ell_i}}{N^\ell_i!} .
\end{equation}
As described in appendix \ref{appendix:gamma_distribution}, combining the two parameters $\bar{N}^\ell$ and $\beta^\ell$ provides a direct sampling procedure for the parameter $\widetilde{N}^\ell$ by simply drawing random variates from a Gamma distribution, which is the conjugate prior of the Poisson distribution, given as:
\begin{equation}
\mathcal{P}\left( \widetilde{N}^\ell \vert \{ N^\ell_i \}, \{ \delta_i \} \right) = \frac{ \left(\widetilde{N}^\ell\right)^{k_\ell-1}\,\mathrm{e}^{-\frac{\widetilde{N}^\ell}{\theta_\ell}}} {\theta^{k_\ell}_\ell \, \Gamma(k_\ell)} ,
\end{equation}
with
\begin{equation}
k_\ell=1+\sum_i N^\ell_i
\end{equation}
and
\begin{equation}
\theta_\ell=\frac{1}{\sum_i R_i^\ell \left( 1+ \delta_i \right)^{\alpha^\ell}}\, .
\end{equation}
\begin{figure*}
\centering{\includegraphics[width=0.5\textwidth,clip=true]{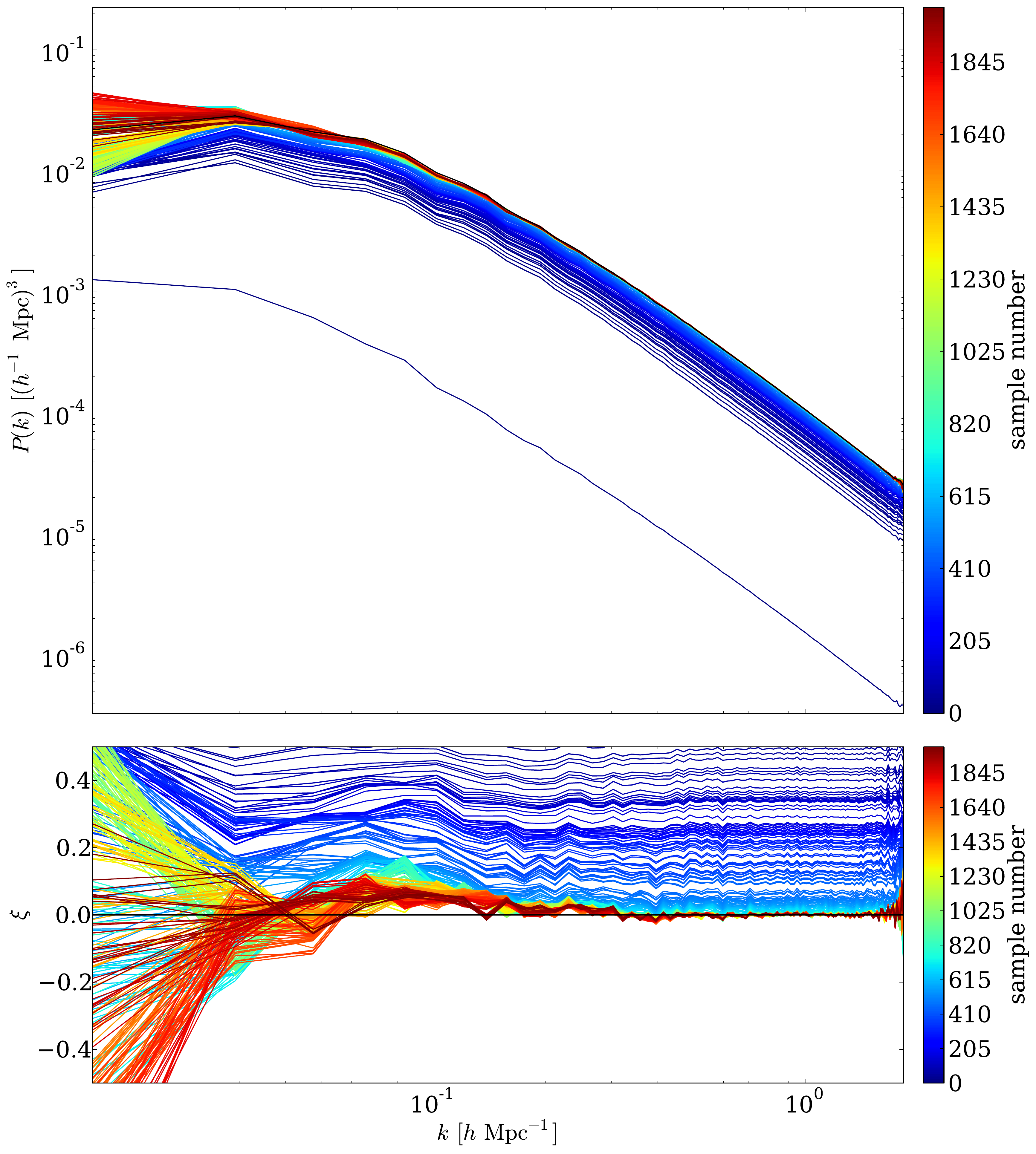}}
\caption{Burn-in power spectra measured from the first 2000 samples of the Markov chain colored corresponding to their sample number as indicated by the colorbar. The black line represents a fiducial reference power spectrum for the cosmology assumed in this work. As can be seen, subsequent power spectra approach the fiducial cosmological power spectrum homogeneously throughout all scales in Fourier space.}
\label{fig:burnin_specs}
\end{figure*}
In a similar fashion as described in \citep{JaschePspec2013}, 3D density fields and noise level parameters can be jointly inferred by adding an additional sampling block to the original formulation of the {\bf{BORG}} algorithm as described in \citep{JASCHEBORG2012}. As indicated by figure \ref{fig:flowchart}, in a first sampling step the algorithm infers 3D initial and final density fields followed by a conditional sampling step to infer the noise level parameters $\widetilde{N}^\ell$. Iteration of this procedure yields Markovian samples from the joint target distribution. For the purpose of this work, we fix the power-law indices $\alpha^\ell$ of the bias relations by requiring them to resemble the linear luminosity dependent bias when expanded in a Taylor series to linear order as:
\begin{equation}
(1+\delta_i)^{\alpha^\ell} = 1 + \alpha^\ell \, \delta_i + \mathcal{O}\left(\delta_i^2\right) . 
\end{equation}
In particular, we assume the functional shape of the luminosity dependent bias parameter $\alpha^\ell$ to follow a standard model for the linear luminosity dependent bias in terms of absolute $r$-band magnitudes $M_{^{0.1}r}$, as given by:
\begin{equation}
\alpha^\ell=b(M_{^{0.1}r}^\ell)=b_{*}\left( a + b \times 10^{0.4\left(M_{*}-M_{^{0.1}r}^\ell \right)} + c \times \left(M_{^{0.1}r}^\ell-M_{*}\right)\right) \, ,
\label{eq:rel_bias}
\end{equation}
with the fitting parameters \(a=0.895\), \(b=0.150\), \(c=-0.040\) and \(M_{*}=-20.40\) \citep[see e.g.][ for details]{2001MNRAS.328...64N,TEGMARK_2004}. The parameter \(b_{*}\) was adjusted during the initial burn-in phase and was finally set to a fixed value of \(b_{*}=1.44\), such that the sampler recovers the correct shape of the assumed initial power spectrum.

\section{The {\bf{BORG}} SDSS analysis}
\label{sec:Bayesian_inference}

The analysis of the SDSS main galaxy sample has been performed on a cubic Cartesian domain with a side length of $750$ Mpc$/h$ consisting of $256^3$ equidistant grid nodes, resulting in $\sim~1.6\times~10^7$ inference parameters. Thus the inference procedure provides data-constrained realizations for initial and final density fields at a grid resolution of about $\sim~3$ Mpc$/h$. For the analysis we assume a standard $\Lambda$CDM cosmology with the set of cosmological parameters $\Omega_m=0.272$, $\Omega_{\Lambda}=0.728$, $\Omega_{b}=0.045$, $h=0.702$, $\sigma_8=0.807$, $n_s=0.961$. The cosmological power spectrum for  initial density fields, has been calculated according to the prescription provided in \citep{1998ApJ...496..605E} and \citep{1999ApJ...511....5E}. In order to sufficiently resolve the final density field, the 2LPT model is evaluated with \(512^3\) particles, by oversampling initial conditions by a factor of eight. 
The entire analysis yielded $12,000$ realizations for initial and final density fields. The generation of a single Markov sample requires a operation count equivalent to about $\sim~200$ 2LPT model evaluations. Typical generation times for data-constrained realizations are shown in the right panel of Figure \ref{fig:markov_properties}. On average the sampler requires about 1500 seconds to generate a single density field realization on \(16\) cores. The total analysis consumed several months of computing time and produced on the order of $\sim~3$ TB of information represented by the set of Markov samples.

The numerical efficiency of any Markov Chain Monte Carlo algorithm, particularly in high dimensions, is crucially determined by the average acceptance rate. As demonstrated by the left panel of Figure \ref{fig:markov_properties}, after an initial burn-in period, the acceptance rate asymptotes at a value of about 60 percent, rendering our analysis numerically feasible. As a simple consistency check, we follow a standard procedure to determine the initial burn-in behavior of the sampler via a simple experiment \citep[see e.g.][for more details]{2004ApJS..155..227E,JASCHE2010HADESMETHOD,JASCHEBORG2012}. The sampler is initialized with an overdispersed state, far remote from the target region in parameter space, by scaling normal random amplitudes of the initial density field at a cosmic scale factor of \(a=10^{-3}\) by a constant factor of 0.01. In the course of the initial burn-in phase, the Markov chain should then drift towards preferred regions in parameter space. As demonstrated by Figure \ref{fig:burnin_specs}, this drift is manifested by a sequence of posterior power spectra measured from subsequent initial density field realizations. It can be clearly seen that the chain approaches the target region  within the first $2000$ sampling steps. The sequence of power spectra shows a homogeneous drift of all modes with no indication of any particular hysteresis or bias across different scales in Fourier space. As improper treatment of survey systematics, uncertainties and galaxy bias typically result in obvious erroneous features in power spectra, Figure \ref{fig:burnin_specs} clearly demonstrates that these effects have been accurately accounted for by the algorithm.

\begin{table}
\center
\renewcommand{\arraystretch}{1.5}
    \begin{tabular}{ccc}
    \hline
      $M_{^{0.1}r}^\ell$ & $\alpha^\ell$ & $\widetilde{N}^\ell$  \\
      \hline
      $-21.00< M_{^{0.1}r}^0 <-20.33$ & 1.58029 & 4.67438 $\times 10^{-2}\, \pm 3.51298\times 10^{-4}$\\
      $-20.33< M_{^{0.1}r}^1 < -19.67$ & 1.41519 & 9.54428 $\times 10^{-2}\, \pm 5.77786\times 10^{-4}$\\
      $  -19.67< M_{^{0.1}r}^2 <-19.00$ & 1.30822 & 1.39989 $\times 10^{-1}\, \pm 1.21087\times 10^{-3}$\\
      $ -19.00< M_{^{0.1}r}^3 <-18.33$ & 1.23272 & 1.74284 $\times 10^{-1}\, \pm 1.89168\times 10^{-3}$\\
      $-18.33< M_{^{0.1}r}^4 <-17.67$  & 1.17424 & 2.19634 $\times 10^{-1}\, \pm 3.42586\times 10^{-3}$\\
      $-17.67 < M_{^{0.1}r}^5 <-17.00$  & 1.12497 & 2.86236 $\times 10^{-1}\, \pm 5.57014\times 10^{-3}$\\
      \hline
    \end{tabular}
     \caption{Bias parameters corresponding to the power-law bias model, as described in the text, for six galaxy sub-samples, subdivided by their absolute $r$-band magnitudes.}
    \label{tb:Table_1}
\end{table}

Note that we also adjust the parameters $\alpha^\ell$ of the assumed power-law bias model during the initial 1000 sampling steps but keep them fixed afterwards. However, as described above, corresponding noise parameters $\widetilde{N}^\ell$ are sampled and explored throughout the entire Markov chain. Inferred ensemble means and standard deviations for the $\widetilde{N}^\ell$ along with chosen power-law parameters $\alpha^\ell$ are provided in Table \ref{tb:Table_1}.

\section{Inference results}
\label{sec:inference_results}

This section describes inference results obtained by our Bayesian analysis of the SDSS main galaxy sample.

\subsection{Inferred 3D density fields}
\label{inf_density_field}

A major goal of this work is to provide inferred 3D initial and final density fields along with corresponding uncertainty quantification in a $\sim~1.6\times 10^7$ dimensional parameter space. To do this, the {\bf{BORG}} algorithm provides a sampled LSS posterior distribution in terms of an ensemble of data-constrained samples, via an efficient implementation of a Markov Chain Monte Carlo algorithm. It should be remarked that, past the initial burn-in phase, all individual samples reflect physically meaningful density fields, limited only by the validity of the employed 2LPT model. In particular, the present analysis correctly accounts for selection effects, survey geometries, luminosity dependent galaxy biases and automatically calibrates the noise levels of the six luminosity bins as described above. As can be seen in Figure \ref{fig:burnin_specs}, past the initial burn-in phase, individual samples possess physically correct power throughout all ranges in Fourier space, and do not show any sign of attenuation due to survey characteristics such as survey geometry, selection effects or galaxy biases.

\begin{figure*}
\centering{\includegraphics[width=1.0\textwidth,clip=true]{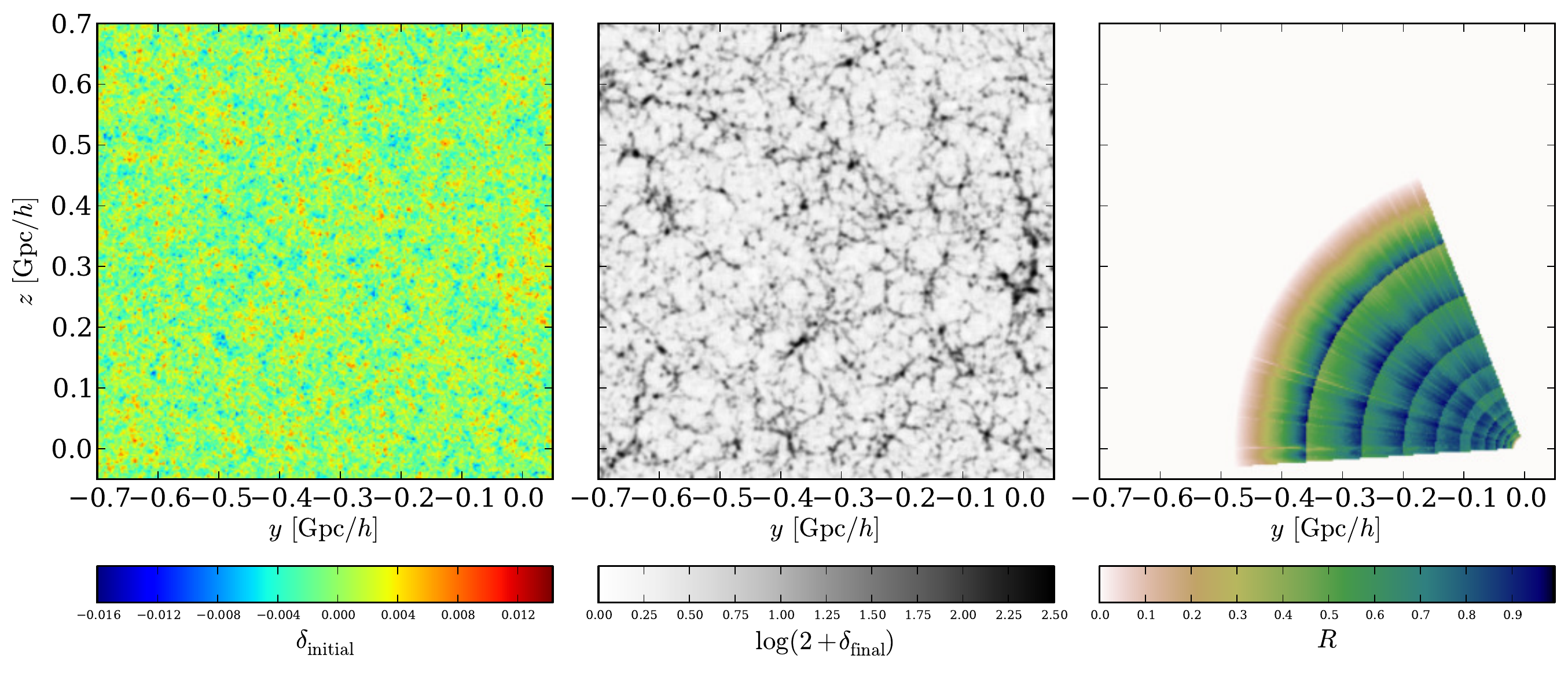}}
\caption{Slices through the initial (left panel) and corresponding final (middle panel) density fields of the $5000$th sample. The right panel shows a corresponding slice through the combined survey response operator $R$ for the six absolute magnitude bins considered in this work.  As can be seen, unobserved and observed regions in the inferred initial and final density fields do not appear visually distinct, demonstrating the fact that individual data-constrained realizations constitute physically meaningful density fields. It also shows that the sampler naturally extends observed large scale structures beyond the survey boundaries in a physically and statistically fully consistent fashion.}
\label{fig:sample_slices}
\end{figure*} 

To further illustrate that individual samples qualify for physically meaningful density fields, in Figure \ref{fig:sample_slices} we show slices through data-constrained realizations of the initial and final density fields of the $5000$th sample as well as corresponding slices through the combined survey response operator $R$, averaged over the six luminosity bins, as discussed above. It can be seen that the algorithm correctly augments unobserved regions with statistically correct information. Note that unobserved and observed regions in the inferred final density fields do not appear visually distinct, a consequence of the excellent approximation of 2LPT not just to the first but also higher-order moments  \citep[][]{MOUTARDE1991,BUCHERT1994,BOUCHET1995,SCOCCIMARRO2000,PTHALOS}. 
Figure \ref{fig:sample_slices} therefore clearly reflects the fact that our sampler naturally extends observed large scale structures beyond the survey boundaries in a physically and statistically fully consistent fashion.
This is a great advantage over previous methods relying on Gaussian or log-normal models specifying the statistics of the density field correctly only to two-point statistics by assuming a cosmological power spectrum. The interested reader may want to qualitatively compare with Figure 2  in \citep{JASCHE2010HADESDATA}, where a log-normal model, unable to represent filamentary structures, was employed.

\begin{figure*}
\centering{\includegraphics[width=1.0\textwidth,clip=true]{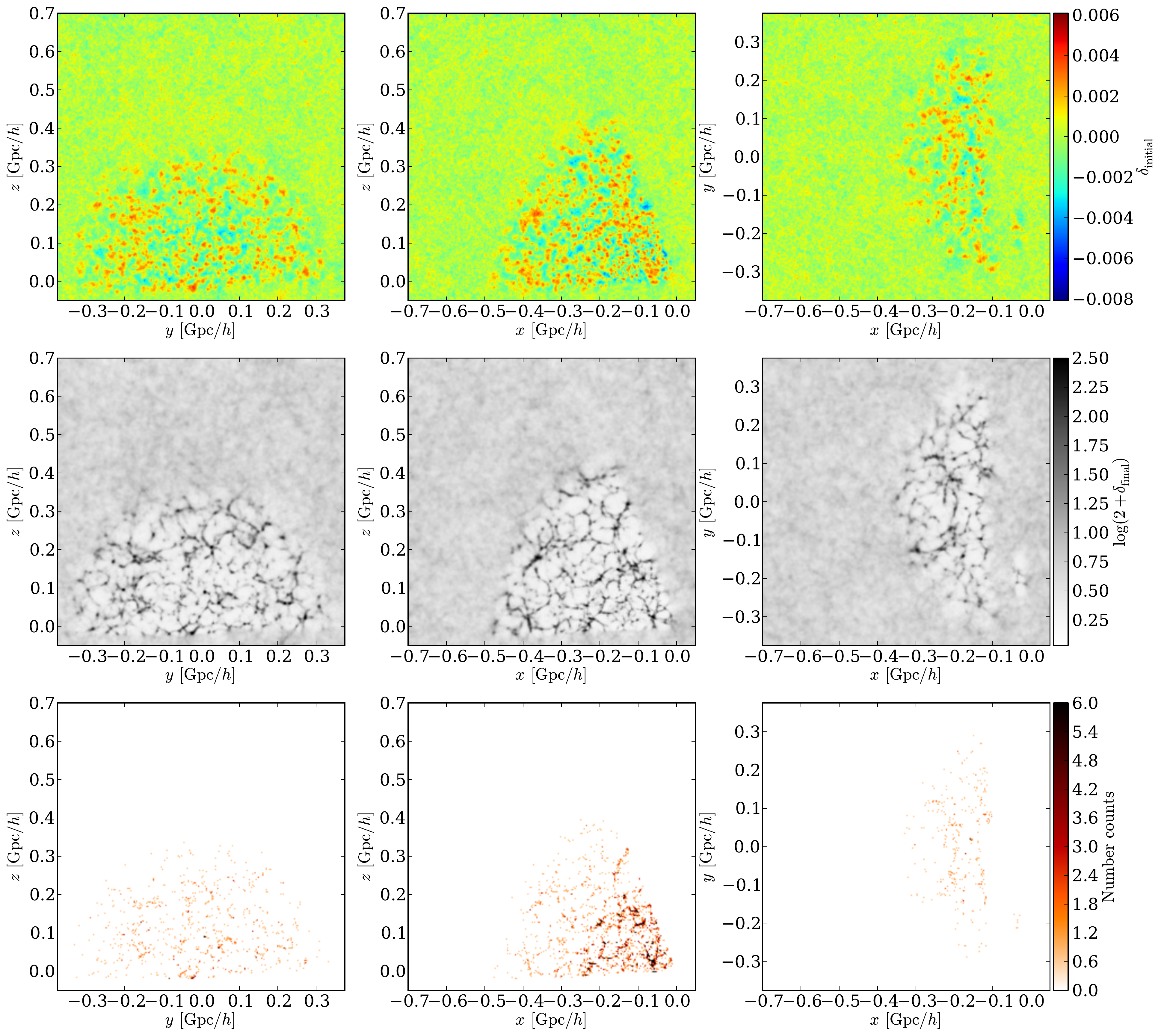}}
\caption{Three slices from different directions through the three dimensional ensemble posterior means for the initial (upper panels) and final density fields (middle panels) estimated from $12,000$ samples. The lower panels depict corresponding slices through the galaxy number counts field of the SDSS main sample.}
\label{fig:mean_var_dens}
\end{figure*}

The ensemble of the $12,000$ inferred data-constrained initial and final density fields permits us to provide any desired statistical summary, such as mean and variance, for full 3D fields. In Figure \ref{fig:mean_var_dens}, we show slices through the ensemble mean initial and final density fields, to be used in subsequent analyses. The plot shows the correct anticipated behavior for inferred posterior mean final density fields, since observed regions represent data constraints, while unobserved regions approach cosmic mean density. This behavior is also present in corresponding initial density fields. In particular, the ensemble mean final density field shows a highly detailed LSS in regions where data constraints are available, and approaches cosmic mean density in regions where data is uninformative on average \citep[see also][for comparison]{JASCHE2010HADESDATA}. Analogously, these results translate to the ensemble mean initial density field. Comparing the ensemble mean final density field to the galaxy number densities, depicted in the lower panels of Figure \ref{fig:mean_var_dens}, demonstrates the performance of the method in regions only poorly sampled by galaxies. In particular, comparing the right middle and right lower panel of Figure \ref{fig:mean_var_dens} reveals the capability of our algorithm to recover highly detailed structures even in noise dominated regions \citep[for a discussion see][]{JASCHEBORG2012}. By comparing ensemble mean initial and final density fields, upper and middle panels in Figure \ref{fig:mean_var_dens}, one can also see correspondences between structures in the present Universe and their origins at a scale factor of $a=10^{-3}$.

\begin{figure*}
\centering{\includegraphics[width=1.0\textwidth,clip=true]{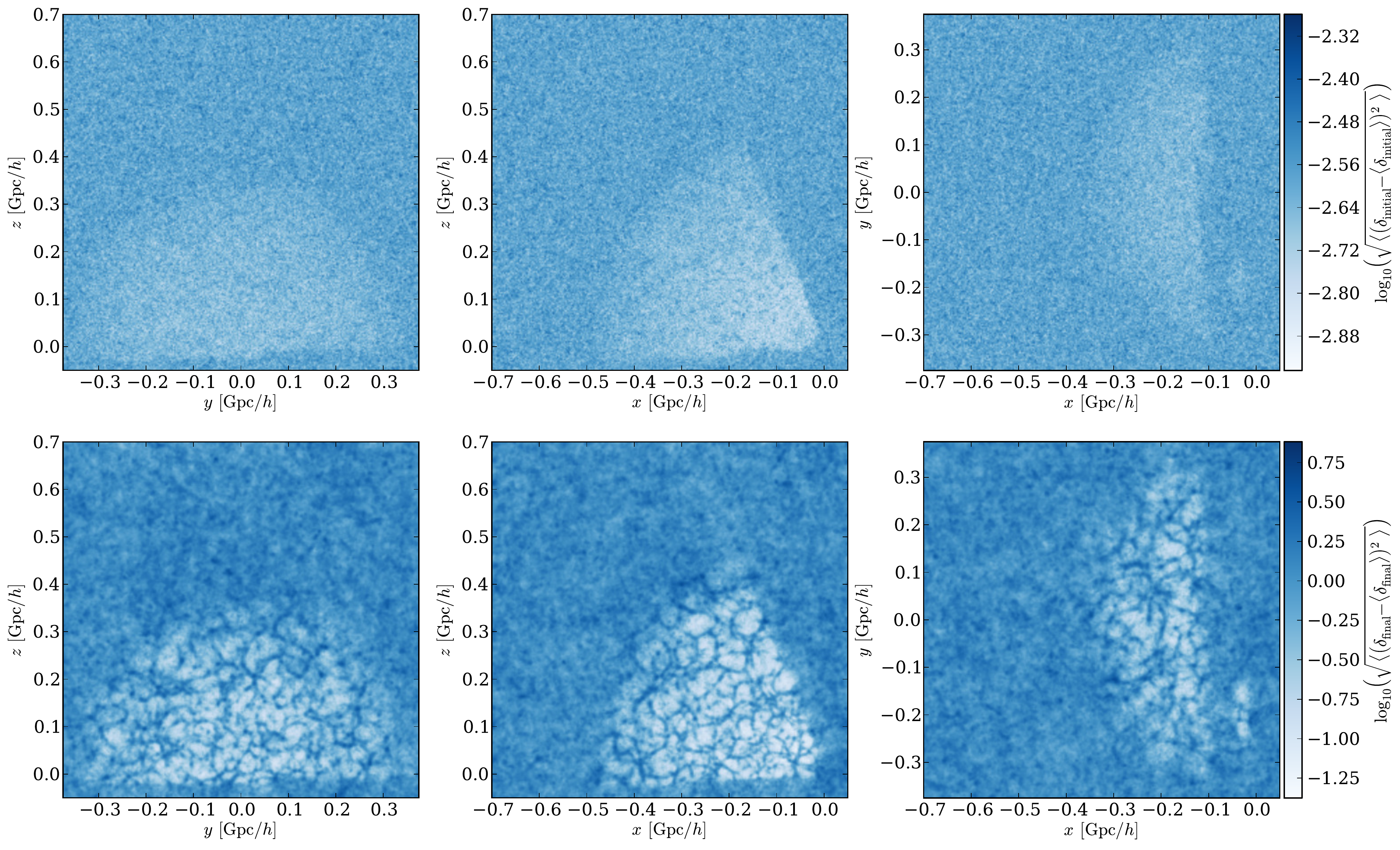}}
\caption{Three slices from different directions through the three dimensional voxel-wise posterior standard deviation for the initial (upper panels) and final density fields (lower panels) estimated from $12,000$ samples. It can be seen that regions covered by observations show on average lower variance than unobserved regions. Also note, that voxel-wise standard deviations for the final density fields are highly structured, reflecting the signal-dependence of the inhomogeneous shot noise of the galaxy distribution. In contrast, voxel-wise standard deviations in the initial conditions are more homogeneously distributed, manifesting the flow of information between data and initial conditions as discussed in the text.}
\label{fig:var_dens}
\end{figure*}

The ensemble of data-constrained realizations also permits to provide corresponding uncertainty quantification. In Figure \ref{fig:var_dens} we plot voxel-wise standard deviations for initial and final density fields estimated from $12,000$ samples. It can be seen that regions covered by data exhibit on average lower variances than unobserved regions, as expected. Note that for non-linear inference problems, signal and noise are typically correlated. This is particularly true for inhomogeneous point processes, such as discrete galaxy distributions tracing an underlying density field. In Figure \ref{fig:var_dens}, the correlation between signal and noise is clearly visible for standard deviation estimates of final density fields. In particular high density regions also correspond to high variance regions, as is expected for Poissonian likelihoods since signal-to-noise ratios scale as the square root of the number of observed galaxies \citep[also see][for a similar discussion]{JASCHE2010HADESDATA}. Also note that voxel-wise standard deviations for final density fields are highly structured, while standard deviations of initial conditions appear to be more homogeneous. This is related to the fact that our algorithm naturally and correctly translates information of the observations non-locally to the initial conditions via Lagrangian transport, as discussed below in section \ref{cosmic_history}.

As mentioned in the introduction, results for the ensemble mean final density field and corresponding voxel-wise standard deviations will be published as as supplementary material to this article. \footnote{Prior to the publication at JCAP please contact us to receive a copy of this data.}

\subsection{Inference of 3D velocity fields}
\label{3d_velocity_fields}

In addition to initial and final density fields, the analysis further provides information on the dynamics of the large scale structure as mediated by the employed 2LPT model. Indeed, the {\bf{BORG}} algorithm shows excellent performance in recovering large scale modes, typically poorly constrained by masked galaxy observations \citep{JASCHEBORG2012}.

\begin{figure*}
\centering{\includegraphics[width=0.65\textwidth,clip=true]{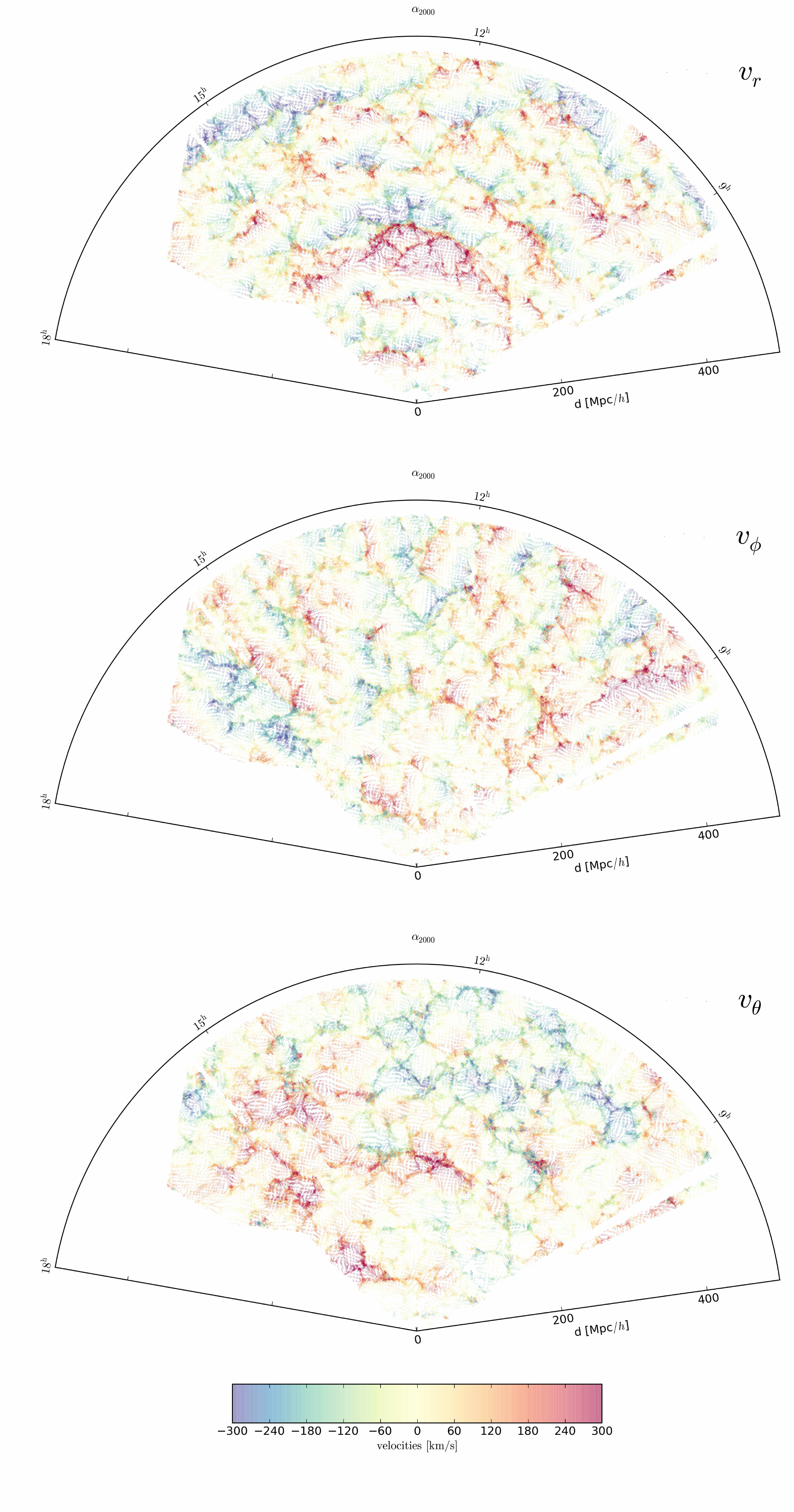}}
\caption{Slices through the 3D velocity fields, derived from the $5000$th sample, for the radial (upper panel), polar (middle panel) and the azimuthal (lower panel) velocity components. The plot shows 2LPT particles in a 4 Mpc$/h$ thick slice around the celestial equator for the observed domain, colored by their respective velocity components.}
\label{fig:velocity_field}
\end{figure*}

This a crucial feature when deriving 3D velocity fields, which are predominantly governed by the largest scales. In this fashion, we can derive 3D velocity fields from our inference results. Note that these velocity fields are derived \textit{a posteriori} and are only predictions of the 2LPT model given inferred initial density fields, since currently the algorithm does not exploit velocity information contained in the data. However, since inferred 2LPT displacement vectors are constrained by observations, and since 2LPT displacement vectors and velocities differ only by constant prefactors given a fixed cosmology, inferred velocities are considered to be accurate. For this reason, exploitation of velocity information contained in the data itself, being subject to a future publication, is not expected to crucially change present results. To demonstrate the capability of recovering 3D velocity fields, in Figure \ref{fig:velocity_field} we show the three components of the velocity field for the $5000$th sample in spherical coordinates. More precisely, Figure \ref{fig:velocity_field} shows the corresponding 2LPT particle distribution evolved to redshift $z=0$  in a $4$ Mpc$/h$ slice around the celestial equator. Particles are colored by their radial (upper panel), polar  (middle panel)  and azimuthal (lower panel) velocity components. To translate between Cartesian and spherical coordinates we used the coordinate transform described in Appendix \ref{appendix:spherical_coordinates}.

\subsection{Inference of LSS formation histories}
\label{cosmic_history}

\begin{figure*}
\centering{\includegraphics[width=1.\textwidth,clip=true]{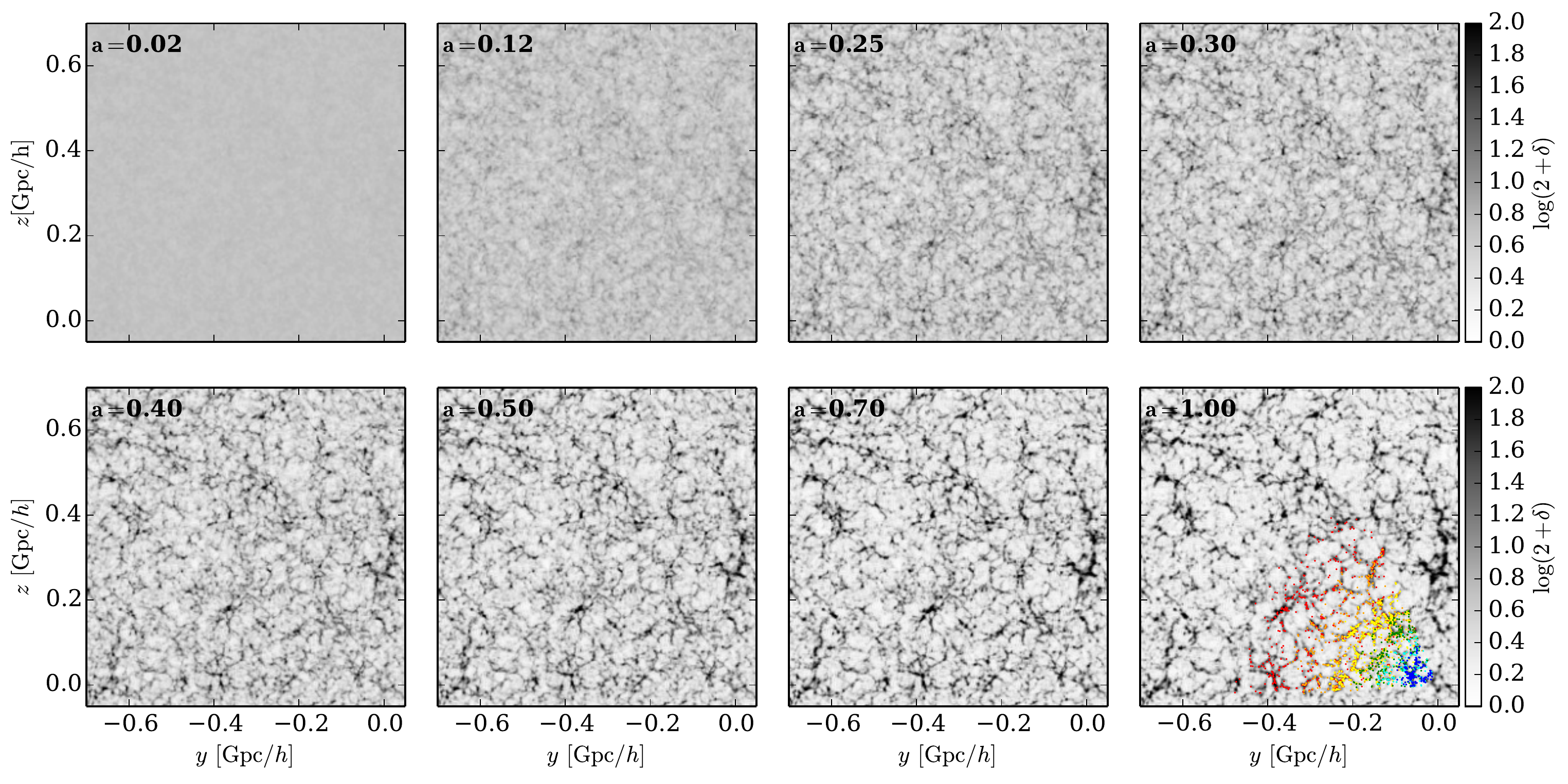}}
\caption{Slices through the inferred three dimensional density field of the $5000$th sample at different stages of its evolution, as indicated by the cosmic scale factor in the respective panels. The plot describes a possible formation scenario for the LSS in the observed domain starting at a scale factor of $a=0.02$ to the present epoch $a=1.0$. In the lower right panel, we overplotted the inferred present density field with the observed galaxies in the respective six absolute magnitude ranges \( -21.00< M_{^{0.1}r} <-20.33\) (red dots), \(-20.33< M_{^{0.1}r} < -19.67\) (orange dots), \(  -19.67< M_{^{0.1}r} <-19.00\) (yellow dots), \( -19.00< M_{^{0.1}r} <-18.33\) (green dots) , \(-18.33< M_{^{0.1}r} <-17.67\) (cyan dots) and , \(-17.67 <M_{^{0.1}r} <-17.00\) (blue dots). As can be clearly seen, observed galaxies trace the recovered three dimensional density field. Besides measurements of three dimensional initial and final density fields, this plot demonstrates that our algorithm also provides plausible four dimensional formation histories, describing the evolution of the presently observed LSS.}
\label{fig:cosmic_evolution}
\end{figure*}

As described above, the {\bf{BORG}} algorithm employs a 2LPT model to connect initial conditions to present SDSS observations in a fully probabilistic approach  \citep{JASCHEBORG2012}. Besides inferred 3D initial and final density fields,  our algorithm therefore also provides full four dimensional formation histories for the observed LSS as mediated by the 2LPT model. As an example, in Figure \ref{fig:cosmic_evolution} we depict the LSS formation history for the 5000th Markov sample ranging from a scale factor of \(a=0.02\) to the present epoch at \(a=1.00\). Initially, the density field seems to obey close to Gaussian statistics and corresponding amplitudes are low. In the course of cosmic history, amplitudes grow and higher-order statistics such as three-point statistics are generated, as indicated by the appearance of filamentary structures. The final panel of Figure \ref{fig:cosmic_evolution}, at a cosmic scale factor of \(a=1.00\), shows the inferred final density field overplotted by SDSS galaxies for the six bins in absolute magnitude, as described previously. Observed galaxies nicely trace the underlying density field. This clearly demonstrates that our algorithm infers plausible formation histories for large scale structures observed by the SDSS survey. By exploring the corresponding LSS posterior distribution, the {\bf{BORG}} algorithm naturally generates an ensemble of such data-constrained LSS formation histories, permitting to accurately quantify the 4D dynamical state of our Universe and corresponding observational uncertainties inherent to galaxy surveys. Detailed and quantitative analysis of these cosmic formation histories will be the subject of forthcoming publications. 

The {\bf{BORG}} algorithm also provides a statistically valid framework for propagating observational systematics and uncertainties from observations to any finally inferred result. This is of particular importance, since detailed treatment of survey geometries and selection effects is a crucial issue if inferred results are to be used for thorough scientific analyses. These effects generally vary greatly across the observed domain and will result in erroneous artifacts if not accounted for properly. Since large scale structure formation is a non-local process, exact information propagation is complex, as it requires to translate uncertainties and systematics from observations to the inferred initial conditions. Consequently, the information content of observed data has to be distributed differently in initial and final density fields, even though the  total amount of information is conserved. Following 2LPT particles from high density regions, and corresponding high signal-to-noise regions in the data, backward in time, demonstrates that the same amount of information contained in the data will be distributed over a larger region in the initial conditions. Analogously, for underdense regions, such as voids, the information content of the data will amass in a smaller volume at the initial state. This means that the signal-to-noise ratio for a given comoving Eulerian volume is a function of time along inferred cosmic histories \citep{JASCHEBORG2012}. This fact manifests itself in the different behaviour of voxel-wise standard deviations for final and initial conditions, as presented in Figure \ref{fig:var_dens}. While the signal-to-noise ratio is highly clustered in final conditions, the same amount of observational information is distributed more evenly over the entire volume in corresponding initial conditions.

\begin{figure*}
\centering{\includegraphics[width=0.8\textwidth,clip=true]{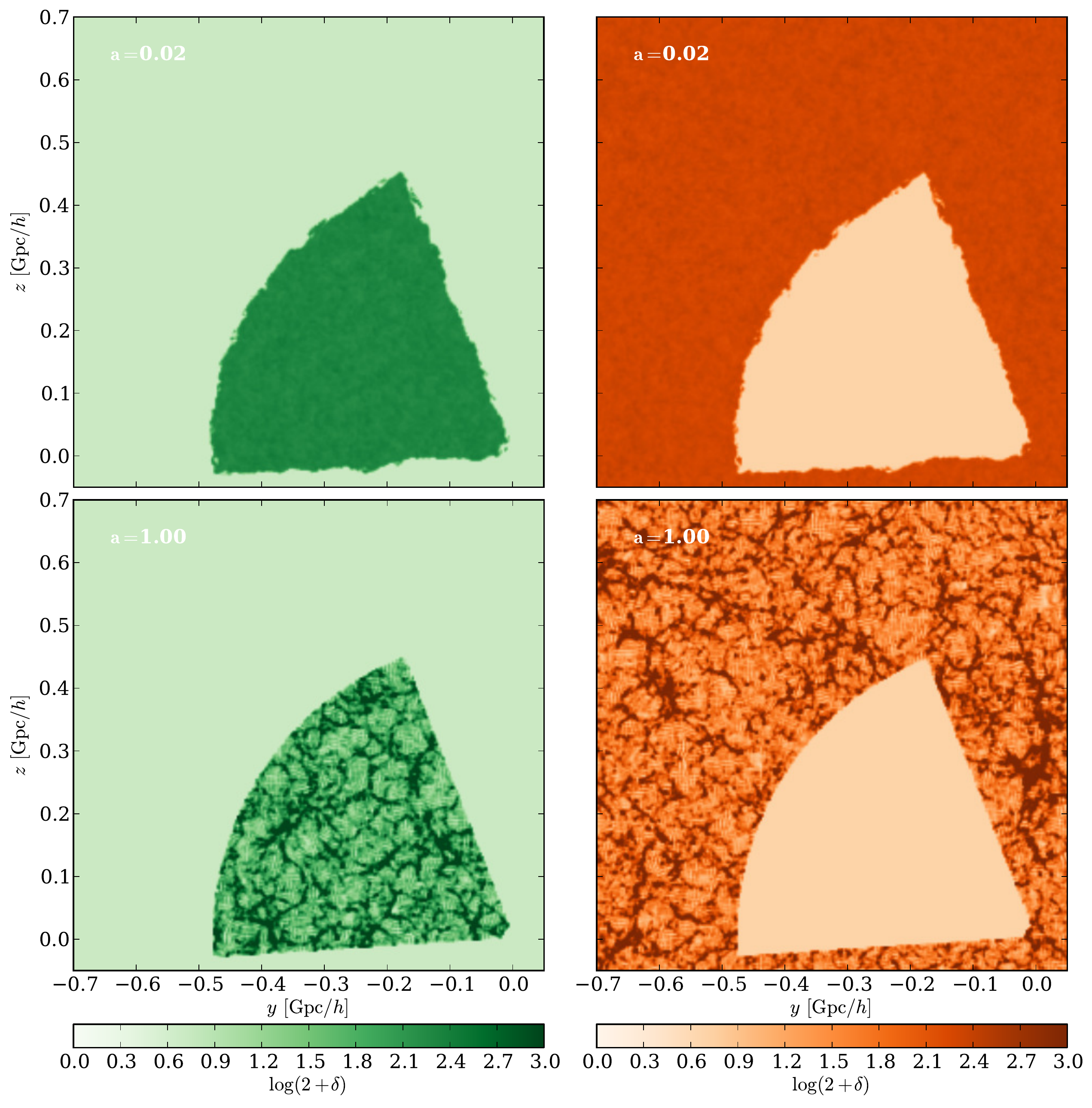}}
\caption{Slices through the distribution of particles in the $5000$th sample, which are located inside (left panels) and outside (right panels) the observed domain at the time of observation, at two time snapshots as indicated in the panels. It can be seen that particles located within the observed region at the present time may originate from regions outside the corresponding comoving Eulerian volume at an earlier epoch and vice versa. As discussed in the text, this plot demonstrates the non-local transport of information, which provides accurate inference of the cosmic large scale structure beyond survey boundaries within a rigorous probabilistic approach.}
\label{fig:inside_out}
\end{figure*}

Non-local propagation of observational information across survey boundaries, together with cosmological correlations in the initial density field, is also the reason why our method is able to extrapolate the cosmic LSS beyond survey boundaries, as discussed in section \ref{inf_density_field} above and demonstrated by Figure \ref{fig:mean_var_dens}. To further demonstrate this fact, in figure \ref{fig:inside_out}, we show the density field of the $5000$th sample traced by particles from inside and outside the observed domain at the present epoch. At the present epoch, the set of particles can be sub-divided into two sets for particles inside and outside the observed domain. The boundary between these two sets of particles is the sharp outline of the SDSS survey geometry. When tracing these particles back to an earlier epoch at a scale factor of $a=0.02$, it can clearly be seen that this sharp boundary starts to frazzle. Particles within the observed domain at the final state may originate from regions outside the corresponding Eulerian volume at the initial state, and vice versa. Information from within the observed domain non-locally influences the large scale structure outside the observed domain, thus increasing the region influenced by data beyond the survey boundaries. Figure \ref{fig:inside_out} therefore demonstrates the ability of our algorithm to correctly account for information propagation via Lagrangian transport within a fully probabilistic approach. The ability to provide 4D dynamic formation histories for SDSS data together with accurate uncertainty quantification paves the path towards high precision chrono-cosmography permitting us to study the inhomogeneous evolution of our Universe. Detailed and quantitative analysis of the various aspects of the results obtained in this work will be subject to future publications.  

\section{Summary and Conclusions}
\label{sec:Summary_an_Conclusion}

This work discusses a fully Bayesian chrono-cosmographic analysis of the 3D cosmological large scale structure underlying the SDSS main galaxy sample \citep[][]{SDSS7}. We presented a data application of the recently proposed {\bf{BORG}} algorithm \citep[for details see][]{JASCHEBORG2012}, which permits to simultaneously infer initial and present non-linear 3D density fields from galaxy observations within a fully probabilistic approach. As discussed in section \ref{sec:methodology}, the algorithm incorporates a second order Lagrangian perturbation model to connect observations to initial conditions and to perform dynamical large scale structure inference from galaxy redshift surveys.

Besides correctly accounting for usual statistical and systematic uncertainties, such as noise, survey geometries and selection effects, this methodology also physically treats gravitational structure formation in the linear and mildly non-linear regime and captures higher order statistics present in non-linear density fields  \citep[see e.g. ][]{MOUTARDE1991,BUCHERT1994,BOUCHET1995,SCOCCIMARRO2000,PTHALOS}. The {\bf{BORG}} algorithm explores a high-dimensional posterior distribution via an efficient implementation of a Hamiltonian Monte Carlo sampler and therefore provides naturally and fully self-consistently accurate uncertainty quantification for any finally inferred quantity \citep[][]{JASCHEBORG2012}. 

In this work, we upgraded the original sampling procedure described in \citep{JASCHEBORG2012} to account for automatic noise calibration and luminosity dependent galaxy biases. To do so, we followed the philosophy described in \citep{JaschePspec2013} and splitted the main galaxy sample into six absolute magnitude bins in the range $-21<M_{^{0.1}r}<-17$. The Bayesian analysis treats each of this six galaxy sub-samples as an individual data set with its individual statistical and systematic uncertainties. As described in section \ref{sec:methodology}, we augmented the original algorithm described in \citep{JASCHEBORG2012} by a power-law bias model and an additional sampling procedure to jointly infer corresponding noise levels for the respective galaxy samples. 

As discussed in section \ref{sec:Bayesian_inference}, we applied this modified version of the {\bf{BORG}} algorithm to the SDSS DR7 main galaxy samples and generated about 12,000 full three dimensional data-constrained initial conditions in the course of this work. The initial density field, at a scale factor of $a=10^{-3}$, has been inferred on a comoving Cartesian equidistant grid, of side length $750$ Mpc$/h$ and $256^3$ grid nodes. This amounts to an target resolution of about $\sim~3~\mathrm{Mpc}/h$ for respective volume elements. Density amplitudes at these Lagrangian grid nodes correspond to about $\sim ~10^7$ parameters to be constrained by our inference procedure. Typically, the generation of individual data-constrained realizations involves an equivalent of $\sim ~200$ 2LPT evaluations and requires on the order of $1500$ seconds on $16$ cores. Despite the complexity of the problem, we demonstrated that our sampler can explore multi-million dimensional parameter spaces via efficient Markov Chain Monte Carlo algorithms with an asymptotic acceptance rate of about 60 percent, rendering our numerical inference framework numerically feasible.

To test the performance of the sampler, we followed a standard approach for testing the initial burn-in behavior via experiments \citep[see e.g.][]{2004ApJS..155..227E,JASCHE2010HADESMETHOD,JASCHEBORG2012}. We initialized the sampler with a Gaussian random field scaled by a factor of $0.01$, to start from an over-dispersed state. During an initial burn-in period the sampler performed a systematic drift towards the target region in parameter space. We examined the initial burn-in behavior by following the sequence of \textit{a posteriori} power spectra, measured from the first 2500 samples, and showed that subsequent samples homogeneously approach the target spectrum throughout all regions in Fourier space without any sign of hysteresis. This indicates the efficiency of the sampler to rapidly explore all scales of the inference problem. The absence of any particular bias or erroneous power throughout all scales in Fourier space, further demonstrates the fact that survey geometry, selection effects, galaxy biasing and observational noise have been accurately accounted for in this analysis. These \textit{a posteriori} power spectra also indicate that individual data-constrained realizations possess the correct physical power in all regions in Fourier space, and can therefore be considered as physically meaningful density fields. This fact has been further demonstrated in section \ref{inf_density_field} by showing slices through an arbitrary data-constrained realization. These results clearly demonstrate the power of our Bayesian methodology to correctly treat the ill-posed inverse problem of inferring signals from incomplete observations, by augmenting unobserved regions with statistically and physically meaningful information. In particular, constrained and unconstrained regions in the samples are visually indistinguishable, demonstrating a major improvement over previous approaches, typically relying on Gaussian or log-normal statistics, incapable of representing the filamentary structure of the cosmic web \citep[see e.g.][]{JASCHE2010HADESDATA}. It should be remarked that this fact not only demonstrates the ability to access high-order statistics in finally inferred quantities such as 3D density maps, but also reflects the control of higher-order statistics in uncertainty quantification far beyond standard normal statistics.

The ensemble of $12,000$ full 3D data-constrained samples permits us to estimate any desired statistical summary. In particular, in section \ref{inf_density_field} we showed ensemble mean density fields for final and initial conditions. A particularly interesting aspect is the fact that the algorithm manages to infer highly-detailed large scale structures even in regimes only poorly covered by observations \citep[for further comments see][]{JASCHEBORG2012}. To demonstrate the possibility of uncertainty quantification, we also calculated the ensemble voxel-wise posterior standard deviation, which reflects the degree of statistical uncertainty at every volume element in the inference domain. As discussed in section \ref{inf_density_field}, these results clearly reflect the signal-dependence of noise for any inhomogeneous point processes, such as discrete Poissonian galaxy distribution. As expected, high signal regions correspond to high variance regions. These results further demonstrate the ability to accurately translate uncertainties in the final conditions to initial density fields, as demonstrated by the plots of voxel-wise standard deviations for corresponding initial density fields. However, note that voxel-wise standard deviations are just a approximation to the full joint and correlated uncertainty that otherwise can by correctly quantified by considering the entire set of data-constrained realizations. Besides 3D initial and final density fields, the methodology also provides information on cosmic dynamics, as mediated by the 2LPT model. In section \ref{3d_velocity_fields}, we showed a velocity field realization in one sample. In particular, we showed the radial, polar and azimuthal velocity components in a 4 Mpc$/h$ thick slice around the celestial equator for the observed domain. These velocities are not primarily constrained by observations, but are derived from the 2LPT model. However, since 2LPT displacement vectors are data-constrained, and since displacement vectors and velocities differ only by constant factors independent of the inference process, derived velocities are considered to be accurate. 

As pointed out frequently, the {\bf{BORG}} algorithm employs 2LPT as a dynamical model to connect initial conditions to present observations of SDSS galaxies. As a consequence, the algorithm not only provides 3D density and velocity fields but also infers plausible 4D formation histories for the observed LSS. In section \ref{cosmic_history}, we illustrated this feature with an individual sample. We followed its cosmic evolution from a initial scale factor of $a=0.02$ to the present epoch at $a=1.00$. As could be seen, the initial density field appears homogeneous and obeys Gaussian statistics. In the course of structure formation clusters, filaments and voids are formed. To demonstrate that this formation history correctly recovers the observed large scale structure, we plotted the observed galaxies, for the six luminosity bins, on top of the final density field. These results clearly demonstrate the ability of our algorithm to infer plausible large scale structure formation histories compatible with observations. Additionally, since the {\bf{BORG}} algorithm is a full Bayesian inference framework, it not only provides a single 4D history, but an ensemble of such data-constrained formation histories and thus accurate means to quantify corresponding observational uncertainties. In particular, our methodology correctly accounts for the non-local transport of observational information between present observations and corresponding inferred initial conditions. As discussed in section \ref{cosmic_history}, the information content in initial and final conditions has to be conserved but can be distributed differently. High-density regions in the final conditions, typically coinciding with high signal-to-noise regions in the data, form by clustering of matter which was originally distributed over a larger Eulerian volume in the initial conditions. For this reason, the observational information associated to a cluster in the final density field will be distributed over a larger volume in the corresponding initial density field. Conversely, the information content of voids in the final conditions will be confined to a smaller volume in the initial conditions. This fact is also reflected by the analysis of voxel-wise standard deviations presented in section \ref{inf_density_field}. While the signal-to-noise ratio is highly clustered in the final conditions, the same amount of observational information is distributed more homogeneously over the entire volume in corresponding initial conditions. As discussed in section \ref{cosmic_history}, particles within the observed domain at the final state may originate from regions outside the corresponding comoving Eulerian volume in the initial conditions and vice versa \citep[also see][]{JASCHEBORG2012}. This non-local translation of information along Lagrangian trajectories is also the reason for the ability of our methodology to extrapolate beyond the survey boundaries of the SDSS and infer the LSS there within a fully probabilistic and rigorous approach. In particular, the high degree of control on statistical uncertainties permit us to perform accurate inferences on the nature of initial conditions and formation histories for the observed LSS in these regions. For these reasons we believe that inferred final ensemble mean fields and corresponding voxel-wise standard deviations as a means of uncertainty quantification, may be of interest to the scientific community. These data products will be published as supplementary material along with this article. \footnote{Prior to the publication at JCAP please contact us to receive a copy of this data.}

In summary, this work describes a application of the previously proposed {\bf{BORG}} algorithm to the SDSS DR7 main galaxy sample. As demonstrated, our methodology produces a rich variety of scientific results, various aspects of which will be objects of detailed and quantitative analyses in forthcoming publications. Besides pure three dimensional reconstructions of the present density field, the algorithm provides detailed information on corresponding initial conditions, large scale dynamics and formation histories for the observed LSS. Together with a thorough quantification of joint and correlated observational uncertainties, these results mark the first steps towards high precision chrono-cosmography, the subject of analyzing the four dimensional state of our Universe. 

\section*{Acknowledgments}
We thank  Guilhem Lavaux, Joseph Silk, Matias Zaldarriaga and Svetlin Tassev for exciting discussions and useful comments. We are also grateful to
Paul M. Sutter, Nico Hamaus and Alice Pisani for their encouragements and support during the realization of this project. Special thanks also go to St\'ephane Rouberol for his support during the course of this work, in particular for guaranteeing flawless use of all required computational resources. This work was partially supported by a Feodor Lynen Fellowship by the Alexander von Humboldt foundation and BW's Chaire d'Excellence from the Agence Nationale de la Recherche (ANR-10-CEXC-004-01) and a Chaire Internationale by UPMC, as well as NSF grants AST 07-08849  and AST 09-08693 ARRA. FL acknowledges funding from an AMX grant (\'Ecole polytechnique ParisTech). This work made in the ILP LABEX (under reference ANR-10-LABX-63) was supported by French state funds managed by the ANR within the Investissements d'Avenir programme under reference ANR-11-IDEX-0004-02.

Funding for the SDSS and SDSS-II has been provided by the Alfred P. Sloan Foundation, the Participating Institutions, the National Science Foundation, the U.S. Department of Energy, the National Aeronautics and Space Administration, the Japanese Monbukagakusho, the Max Planck Society, and the Higher Education Funding Council for England. The SDSS Web Site is http://www.sdss.org/.

The SDSS is managed by the Astrophysical Research Consortium for the Participating Institutions. The Participating Institutions are the American Museum of Natural History, Astrophysical Institute Potsdam, University of Basel, University of Cambridge, Case Western Reserve University, University of Chicago, Drexel University, Fermilab, the Institute for Advanced Study, the Japan Participation Group, Johns Hopkins University, the Joint Institute for Nuclear Astrophysics, the Kavli Institute for Particle Astrophysics and Cosmology, the Korean Scientist Group, the Chinese Academy of Sciences (LAMOST), Los Alamos National Laboratory, the Max-Planck-Institute for Astronomy (MPIA), the Max-Planck-Institute for Astrophysics (MPA), New Mexico State University, Ohio State University, University of Pittsburgh, University of Portsmouth, Princeton University, the United States Naval Observatory, and the University of Washington.

\bibliographystyle{plain}
\include{biblio}

\appendix

\section{The $\Gamma$-distribution for noise sampling}
\label{appendix:gamma_distribution}

As described in the main text, we aim at automatically calibrating the noise levels of individual galaxy samples
within the inference process. Following the philosophy of \citep{JaschePspec2013}, this can be achieved by introducing an additional sampling block to the original implementation of the  {\bf{BORG}} algorithm as presented in \citep{JASCHEBORG2012}. This additional sampling block is designed to provide random variates of the noise parameter $\widetilde{N}^\ell$ conditioned by the galaxy data $\{ N^\ell_i \}$ and the current final density sample $\{\delta_i\}$. According to the Bayes formula, we can write:
\begin{equation}
\label{appendix:eq_noise_level_post}
\mathcal{P}\left( \widetilde{N}^\ell \vert \{ N^\ell_i\},\{\delta_i\} \right) \propto \mathcal{P}\left( \widetilde{N}^\ell\right) \, \mathcal{P}\left( \{ N^\ell_i\} \vert \widetilde{N}^\ell,\{\delta_i\}\right) \, , 
\end{equation}
where we assumed the conditional independence $\mathcal{P}\left( \{\delta_i\} \vert \widetilde{N}^\ell \right) = \mathcal{P}\left( \{\delta_i\} \right)$. In the absence of any further information on the parameter $\widetilde{N}^\ell$, we follow the maximum agnostic approach pursued by Jasche and Wandelt in\citep{JaschePspec2013} by setting the prior distribution $\mathcal{P}\left( \widetilde{N}^\ell\right)$ constant.
By introducing the Poisson likelihood $\mathcal{P}\left( \{ N^\ell_i\} \vert \widetilde{N}^\ell,\{\delta_i\}\right)$ described in equation (\ref{eq:poisson_likelihood_2}) into equation (\ref{appendix:eq_noise_level_post}) we obtain the conditional posterior for the noise parameter $\widetilde{N}^\ell$ as:
\begin{equation}
\mathcal{P}\left( \widetilde{N}^\ell \vert \{ N^\ell_i\},\{\delta_i\} \right) \propto \mathrm{e}^{-\widetilde{N}^\ell A_\ell} \, \left(\widetilde{N}^\ell\right)^{B_\ell} ,
\end{equation}
where $A_\ell \equiv \sum_i R_i^\ell \left( 1+\delta_i \right)^{\alpha^\ell}$ and $B_\ell \equiv \sum_i N^\ell_i$. 
By choosing $k_\ell \equiv B_\ell+1$ and $\theta_\ell \equiv 1/A_\ell$ we yield a properly normalized Gamma distribution for the noise parameter $\widetilde{N}^\ell$, given as:
\begin{equation}
\mathcal{P}\left( \widetilde{N}^\ell \vert \{ N^\ell_i\},\{\delta_i\} \right) = \frac{ \left(\widetilde{N}^\ell\right)^{k_\ell-1}\,\mathrm{e}^{-\frac{\widetilde{N}^\ell}{\theta_\ell}}} {\theta^{k_\ell}_\ell \, \Gamma(k_\ell)} .
\end{equation}
Random variates of the Gamma distribution can be generated by standard sampling routines, such as provided by the GNU scientific library \citep[][]{GSL}.

\section{Spherical to Cartesian coordinates}
\label{appendix:spherical_coordinates}
To translate between spherical and Cartesian coordinates we used the following
coordinate transform:
\begin{eqnarray}
x&=& d_{com}\,cos(\lambda)\,cos(\eta)\nonumber\\
y&=& d_{com}\,cos(\lambda)\,sin(\eta)\\
z&=& d_{com}\,sin(\lambda)\nonumber
\end{eqnarray}
with \(\lambda\) being the declination, \(\eta\) being the right ascension and $d_{com}$ being the radial co-moving distance.

\label{lastpage}

\end{document}

%% file: biblio.tex
\bibliography{paper}
\bibliographystyle{mn2e}